\documentclass[a4paper,10pt,aps,amsfonts,amsmath,nofootinbib,byrevtex,prd,notitlepage]{revtex4-1}

\usepackage[UKenglish]{babel}
\usepackage{drcmath}
\DeclareVector{\valpha}{\alpha}

\usepackage{array,booktabs}
\usepackage[utf8]{inputenc}

\begin{document}

\title{Dynamical Quark Mass Generation in QCD\textsubscript{3} within the Hamiltonian approach in Coulomb gauge}

\author{Felix Spengler}
\author{Davide Campagnari}
\author{Hugo Reinhardt}
\affiliation{Institut f\"ur Theoretische Physik, Universit\"at T\"ubingen,
Auf der Morgenstelle 14, 72076 T\"ubingen, Germany}
\date{\today}


\begin{abstract}
We investigate the equal-time (static) quark propagator in Coulomb gauge within
the Hamiltonian approach to QCD in $d=2$ spatial dimensions. Although the underlying
Clifford algebra is very different from its counterpart in $d=3$, the gap equation
for the dynamical mass function has the same form. The additional vector kernel
which was introduced in $d=3$ to cancel the linear divergence of the gap equation
and to preserve multiplicative renormalizability of the quark propagator makes the
gap equation free of divergences also in $d=2$.
\end{abstract}

\maketitle


\section{Introduction}

The two most striking features of low-energy Quantum Chromodynamics (QCD) at ordinary
temperature and density are confinement and the spontaneous breaking of chiral symmetry.
In recent years the research interest has been shifted to the investigation of thermal
properties of QCD and of its phase diagram, where a central challenge is to locate
the critical end point.
Experimentally, there has been progress at the Relativistic
Heavy-Ion Collider and the Large Hadron Collider; searches for the critical end point
are on-going at the FAIR and NICA facilities. On the theoretical side,
lattice calculations are hindered by the notorious sign problem at finite chemical
potential. Furthermore, simulating three or four families of light dynamical quarks
involves a high computational cost; an approach to reducing this cost is to reduce
the number of physical dimensions. 

QCD in $1+1$ dimensions has been widely studied as toy model and in fact displays
some relevant properties of real QCD, but fails to be a reliable testing ground for QCD\textsubscript{4}, since
gauge symmetries are somewhat trivial in two dimensions (unless a compact manifold
is considered). QCD in $2+1$ dimensions is a more interesting alternative, which
moreover allows the addition of a topological Chern--Simons term.

In this paper we examine QCD\textsubscript{3} with one massless fermion within the Hamiltonian
approach in Coulomb gauge developed previously in $d=3$ spatial dimensions \cite{Feuchter:2004mk,Pak:2013uba,Vastag:2015qjd}.
Within this approach we will investigate how a
mass is dynamically generated by the interaction with the gluons. 
Our previous work in $d=3$ \cite{Pak:2011wu,Pak:2013uba,Vastag:2015qjd,Campagnari:2016wlt,Campagnari:2018flz}
has shown that the linearly rising colour Coulomb potential is the trigger of
chiral symmetry breaking, and that a genuinely non-perturbative Dirac structure in the quark-gluon vertex
eliminates the linear divergence of the quark gap equation and makes the latter ultraviolet (UV) finite.
In the present paper we investigate
whether this cancellation of the UV divergences in the quark gap equation persists also in $d=2$.
This is by far not obvious since the algebra of the Dirac matrices is different
in $d=2$ from the $d=3$ case and moreover the degree of divergence is different.
Of course, quarks in $d=2$ have no chiral symmetry to be broken\footnote{%
   Although with an even number of fermion fields one can mimic chiral symmetry~\cite{Pisarski:1984dj},
   and the corresponding ``chiral symmetry breaking'' is in fact a flavour symmetry breaking}
since there is no counterpart of $\gamma_5$
in $d=2$. The interesting question is here how a dynamical quark mass, which in $d=3$ is a
consequence of spontaneous breaking of chiral symmetry, is generated in $d=2$ without chiral symmetry breaking.
We will show that within the Hamiltonian approach to QCD in Coulomb gauge the dynamical quark mass generation
is caused in $d=2$ by the confining non-abelian Coulomb interaction of the quarks, like in the $d=3$ case.
So within this approach the dynamical mass generation seems to be a universal phenomenon which is
independent of the number of dimensions and not necessarily linked to the spontaneous breaking of
chiral symmetry. We will also show that the cancellation of the leading UV divergence in the quark gap equation
found in $d=3$ with our Ansatz for the quark vacuum wave functional occurs in \emph{any} dimension.

The structure of the paper is as follows: In Sec.~\ref{sec:qcd3} we review the Hamiltonian
approach to QCD with the modifications for $d=2$; in Sec.~\ref{sec:ans} we present our Ansatz for the QCD vacuum
wave functional and show that in the bare-vertex approximation, where the full quark-gluon vertex
is replaced by the bare one, the quark propagator satisfies
the same Dyson--Schwinger equation (DSE) known from $d=3$ in \emph{any} number of dimensions;
in Sec.~\ref{sec:en} we present the evaluation of the energy density and derive the gap equations
for the variational kernels occurring in the Ansatz for the vacuum wave functional. The numerical
results are presented in Sec.~\ref{sec:num} and our conclusions are given in Sec~~\ref{sec:conc}.


\section{QCD in Two Space Dimensions}
\label{sec:qcd3}
The Hamilton operator of QCD in Coulomb gauge $\divg\vA=0$ reads \cite{Christ:1980ku}
\begin{equation}\label{hh1}
\begin{split}
H_{\mathrm{QCD}} ={}& \frac12 \int\d^2x \, J_A^{-1} \Pi_i^a(\vx) J_A \, \Pi_i^a(\vx) + \frac12 \int\d^3x \, B_i^a(\vx) \, B_i^a(\vx) \\
     &+ \int\d^2x \, \psi^\dag(\vx) \bigl[ -\I*\valpha\cdot\grad - g \valpha\cdot\vec{A}(\vx) + \beta m \bigr] \psi(\vx) + H_{\mathrm{C}},
\end{split}
\end{equation}
where $\vec{A}=\vec{A}^{\!a\,} t^a$ are the (transverse) spatial gauge fields
with $t^a$ being the hermitian generators of the $\mathfrak{su}(\Nc)$ algebra,
$\Pi_i^a=-\I*\delta/\delta A_i^a$ is the canonical momentum, $B_i^a$ is the chromomagnetic
field, and $J_A=\Det G_A^{-1}$  is the Faddeev--Popov determinant. Furthermore,
$\psi$ and $\psi^\dag$ are the quark field operators, $\alpha_i$ and $\beta$
are the Dirac matrices (which in $d=2$ coincide with Pauli matrices), and~$m$
is the bare current quark mass. The last term in \Eqref{hh1} is the so-called Coulomb term
\[
H_{\mathrm{C}} = \frac{g^2}{2} \int \d^2x \d^2y \, J_A^{-1} \rho^a(\vx) \,J_A \, F_A^{ab}(\vx,\vy) \, \rho^b(\vy) ,
\]
which describes the interaction of the colour charge density
\[
\rho^a(\vx) = \psi^\dag(\vx) \, t^a \psi(\vx) + f^{abc} A_i^b(\vx) \, \Pi_i^c(\vx)
\]
through the Coulomb kernel
\begin{equation}\label{coulkernel}
F_A^{ab}(\vx,\vy) = \int\d^3z \, G_A^{ac}(\vx,\vz) \bigl( - \nabla^2_z \bigr) G_A^{cb}(\vz,\vy) ,
\end{equation}
where
\[
G_A^{-1}(\vx,\vy) = \bigl( - \delta^{ab} \nabla^2_x - g f^{acb} A_i^c(\vx) \partial_i^x \bigr) \delta(\vx-\vy)
\]
is the Faddeev--Popov operator of Coulomb gauge with $f^{acb}$ being the structure constants
of the $\mathfrak{su}(\Nc)$ algebra.

For the Dirac matrices we choose the ``Dirac'' representation where $\beta$ is
diagonal
\[
\alpha_{i=1,2} = \sigma_{i=1,2} , \qquad \beta = \sigma_3 .
\]
They satisfy the usual Dirac algebra
\[
\anticomm{\alpha_1}{\alpha_j} = \delta_{ij} , \qquad \anticomm{\alpha_i}{\beta} = 0, \qquad \beta^2=1 .
\]
In two dimensions we have
\[
\comm{\alpha_i}{\alpha_j} = 2 \mkern2mu \I*\eps_{ij} \beta
\]
which leads to
\[
\alpha_i \alpha_j = \delta_{ij} + \I*\eps_{ij} \beta , \qquad \beta\alpha_i = \I*\eps_{ij} \alpha_j .
\]
For comparison, in $d=3$ we have
\[
\alpha_i \alpha_j = \delta_{ij} + \I*\eps_{ijk} \gamma_5 \alpha_k \qquad (d=3).
\]
The crucial difference to $d=3$ spatial dimensions is that in $d=2$ there is no $\gamma_5$
and accordingly no chiral symmetry.


\section{Vacuum wave functional and quark propagator}
\label{sec:ans}
In the variational approach developed in
Refs.~\cite{Feuchter:2004mk,Epple:2006hv,Campagnari:2010wc,Campagnari:2015zsa,Campagnari:2018flz}
one attempts to solve the functional Schrödinger equation
\[
H_{\mathrm{QCD}} \ket{\varPsi} = E \ket{\varPsi}
\]
for the QCD vacuum state $\ket{\varPsi}$ by means of the variational principle
with suitable trial Ansätze for the vacuum wave functional $\varPsi[A]=\braket{A}{\varPsi}$.
Inspired by the form of the QCD Hamiltonian the
vacuum state is assumed of the form
\begin{equation}\label{evg0}
\ket{\varPsi} = \ket{\varPsi_{\mathrm{YM}}} \ket{\varPsi_{\mathrm{Q}}}
\end{equation}
where $\ket{\varPsi_{\mathrm{YM}}}$ is the vacuum state of the Yang--Mills sector
and $\ket{\varPsi_{\mathrm{Q}}}$ is the vacuum state of the quark sector, which
includes also the coupling to the gluons. Furthermore it turns out that it is
most convenient to use the coordinate representation $\varPsi[A,\xi] = \braket{A,\xi}{\varPsi}$
where $A$ are classical gauge fields and $\xi$ are Grassmann variables, the ``classical
coordinates'' of the fermions. In accordance with \Eqref{evg0} we
write the vacuum wave functional of QCD in the form
\begin{equation}\label{evg2}
\varPsi[A,\xi^\dag_+,\xi_-] \propto \exp\biggl\{ -\frac12 S_A[A] - S_f [\xi_+^\dag,\xi_-,A] \biggr\} ,
\end{equation}
where $S_A$ and $S_f$ define respectively the wave functional of pure Yang--Mills theory
and of the quarks interacting with the gluons. For $S_A$ we could
take a Gaussian Ansatz \cite{Feuchter:2004mk} or a more general form involving
cubic and quartic couplings \cite{Campagnari:2010wc}. However, in the present
paper we do not solve the gluon gap equation but use instead for the gluon propagator
a form which is inspired by the IR an UV analysis of the variational equations
(which besides the gluon gap equations consist also of Dyson--Schwinger-type of
equations, see Ref.~\cite{Campagnari:2010wc}) and which fits the lattice data,
see \Eqref{gribov} below. For $S_f$ we make the Ansatz used already in $d=3$ \cite{Vastag:2015qjd}
\begin{equation}\label{ans2}
S_f [\xi_+^\dag,\xi_-,A] = \int \xi_+^\dag \bigl[ \beta s + g (V + \beta W) \valpha\cdot \vA \bigr] \xi_-
= \int \xi^\dag \Lambda_+ \bigl[ \beta s + g (V + \beta W) \valpha\cdot \vA \bigr] \Lambda_- \xi,
\end{equation}
where $s$, $V$, and $W$ are variational kernels which will determined by the minimization
of the energy density. Due to the coupling of the quarks to the gluons contained
in $S_f$ the wave functional \Eqref{ans2} is necessarily non-Gaussian.
The vacuum expectation value of an operator $O$ is given by the functional integral
\cite{Campagnari:2015zsa,Campagnari:2018flz}
\begin{equation}\label{evg1}
\begin{aligned}[b]
\bra{\varPsi} O\bigl[A,\Pi,\psi,\psi^\dag\bigr] \ket{\varPsi}
= {}&{} \int \calD \xi \calD \xi^\dag \calD A \, J_A \e^{-\mu} \, \varPsi^*[\xi_+^\dag,\xi_-,A] \\
&\times O \biggl[A,-\I \frac{\delta}{\delta A}, \xi_-+\frac{\delta}{\delta\xi_+^\dag}, \xi_+^\dag+\frac{\delta}{\delta\xi_-} \biggr]
\varPsi[\xi_+^\dag,\xi_-,A] ,
\end{aligned}
\end{equation}
where
\[
\mu = \Lambda_+ - \Lambda_-
\]
is the integration measure of the coherent fermion states, and
\[
\xi_\pm(\vx) = \int \d^2 y \, \Lambda_\pm(\vx,\vy) \, \xi(\vy)
\]
are spinor-valued Grassmann fields, with
\begin{equation}\label{proj}
\Lambda_\pm(\vx,\vy) = \int \dfr[2]{p} \e^{\I*\vp\cdot(\vx-\vy)} \biggl( \frac{1}{2} \pm \frac{\valpha\cdot\vp + \beta m}{2\sqrt{\vp^2+m^2}}\biggr)
\end{equation}
being the projectors onto the positive/negative eigenstates of the free Dirac Hamilton operator $\valpha\cdot\vp+\beta m$.

When the form \Eqref{evg2} of the wave functional is inserted into \Eqref{evg1}
and the functional derivatives are worked out, the expectation value of an operator
reduces to a quantum average of field functionals reminiscent of a Euclidean
field theory with action
\[
S = S_A + S_f^{} + S_f^* + \mu .
\]
This formal equivalence can be exploited to derive Dyson--Schwinger-like
equations \cite{Campagnari:2010wc,Campagnari:2015zsa} to express the various Green's
functions in terms of the variational kernels contained in the non-Gaussian ``action''~S.

The essential quantity of the quark sector is the two-point correlation function
of the Grassmann fields $\xi$
\[
Q(\vx,\vy) = \vev{\xi(\vx) \, \xi^\dag(\vy)} ,
\]
which can be parametrized in momentum space as
\[
Q^{-1}(\vp) = A(\vp) \, \valpha\cdot\vp + B(\vp) \, \beta .
\]
With our Ansatz [Eqs.~\eqref{evg2} and \eqref{ans2}] for the vacuum wave functional
the dressing functions $A$ and $B$ obey the Dyson--Schwinger-like equations \cite{Campagnari:2018flz}
\begin{subequations}\label{qdse1}
\begin{align}
\label{qdse1a}
A(\vp) &= 1 - \frac{C_F}{2} \int \dfr[2]{q} \tr\bigl[ \valpha\cdot\uvp \, \bar\Gamma_{0,i}(\vp,-\vq) \, Q(\vq) \, D_{ij}(\vp-\vq) \, \bar\Gamma_j(\vq,-\vp) \bigr] , \\
\label{qdse1b}
B(\vp) &= s(\vp) - \frac{C_F}{2} \int \dfr[2]{q} \tr\bigl[ \beta \, \bar\Gamma_{0,i}(\vp,-\vq) \, Q(\vq) \, D_{ij}(\vp-\vq) \, \bar\Gamma_j(\vq,-\vp) \bigr] ,
\end{align}
\end{subequations}
where $C_F=(\Nc^2-1)/(2\Nc)$ is the quadratic Casimir in the fundamental representation, and
\begin{equation}\label{gluonprop}
\delta^{ab} D_{ij}(\vx,\vy) = \vev{A_i^a(\vx) \, A_j^b(\vy)} , \qquad D_{ij}(\vp) = \frac{t_{ij}(\vp)}{2\Omega(\vp)} ,\qquad
t_{ij}(\vp) = \delta_{ij} - \frac{p_i \, p_j}{\vp^2}
\end{equation}
is the gluon propagator. Furthermore, $\bar\Gamma$ is the full quark-gluon vertex
defined by \cite{Campagnari:2015zsa}
\begin{equation}\label{fqgv}
\vev{\xi \, \xi^\dag \, A_i} = - Q \, \bar{\Gamma}_j \, Q \, D_{ji}
\end{equation}
while $\bar\Gamma_0$ is the bare quark-gluon vertex defined by our Ansatz \Eqref{ans2} for
the vacuum wave functional
\begin{equation}\label{bqgv}
\bar\Gamma_{0,i}(\vp,\vq) = \Lambda_+(\vp) K_i \Lambda_-(-\vq) + \Lambda_-(\vp) K_i^\dag \Lambda_+(-\vq)
\end{equation}
where
\[
K_i = g \bigl[ V(\vp,\vq) + \beta \, W(\vp,\vq) \bigr] \alpha_i .
\]
In the bare-vertex approximation, where the full quark-gluon
vertex \Eqref{fqgv} is replaced by the bare one \Eqref{bqgv}, Eqs.~\eqref{qdse1}
become in the chiral limit $m=0$
\begin{subequations}\label{qdse2}
\begin{align}
\label{qdse2a}
A_p &= 1 - \frac{g^2 C_F}{2} \int \dfr[2]{q} \frac{A_q}{A_q^2+B_q^2} \frac{X_-(\vp,\vq) \, V^2(\vp,\vq) +  X_+(\vp,\vq) \, W^2(\vp,\vq)}{\Omega(\vp+\vq)} , \\
\label{qdse2b}
B_p &= s_p - \frac{g^2 C_F}{2} \int \dfr[2]{q}  \frac{B_q}{A_q^2+B_q^2} \frac{X_-(\vp,\vq) \, V^2(\vp,\vq) -  X_+(\vp,\vq) \, W^2(\vp,\vq)}{\Omega(\vp+\vq)} ,
\end{align}
\end{subequations}
where we have defined the momentum overlap functions
\begin{equation}\label{xpm}
X_\pm(\vp,\vq) = \frac{1\mp\uvp\cdot\uvq}{2} \pm \frac{[\uvp\cdot(\vp+\vq)] [\uvq\cdot(\vp+\vq)]}{(\vp+\vq)^2} .
\end{equation}
In order to simplify the notation, we have denoted in \Eqref{qdse2} the momentum dependence of
the dressing functions by a subscript. As we will show now, these equations hold
formally in any dimension. Only the overlap functions $X_\pm$ depend on the number~$d$
of spatial dimensions. Although the Dirac matrices depend on the number of
spatial dimensions,
when the bare vertices are contracted with the (symmetric) transverse projectors
only anti-commutators, which are independent of the number of dimensions, enter
the final result.
To see this, we write the bare quark-gluon vertex \Eqref{bqgv} in the chiral
limit as
\[
\bar\Gamma_{0,i}(\vp,\vq) = \frac14 \, V(\vp,\vq) \bigl[ M_i(\vp,\vq) + M_i(-\vp,-\vq) \bigr]
- \frac14 \, W(\vp,\vq) \bigl[ M_i(\vp,-\vq) - M_i(-\vp,\vq) \bigr] \beta
\]
where
\[
M_i(\vp,\vq) = (1+\valpha\cdot\uvp) \alpha_i (1+\valpha\cdot\uvq) .
\]
This quantity has the properties
\[
\valpha\cdot\uvp \, M_i(\vp,\vq) = M_i(\vp,\vq) = M_i(\vp,\vq) \, \valpha\cdot\uvq, \qquad \beta M_i(\vp,\vq) = - M_i(-\vp,-\vq) \beta .
\]
Using these relations in the DSEs \eqref{qdse1} leads then to terms of the form
\[
M_i(\pm\vp,\vq) \, M_j(\vq,\pm\vp) .
\]
For general indices $i$, $j$ this will be in general a complicated expression
whose details depend on the number of dimensions. However, in DSEs \eqref{qdse1}
these expressions are always contracted with a transverse projector~$t_{ij}$
[stemming from the gluon propagator \Eqref{gluonprop}], resulting in
\[
t_{ij}(\vp+\vq) \, M_i(\pm\vp,\vq) \, M_j(\vq,\pm\vp)
= 8 (1\pm\valpha\cdot\uvp) X_{\mp}^{(d)}(\vp,\vq)
\]
with
\[
X_{\pm}^{(d)}(\vp,\vq) = \frac{d-1\mp(3-d)\uvp\cdot\uvq}{2} \pm \frac{[\uvp\cdot(\vp+\vq)] [\uvq\cdot(\vp+\vq)]}{(\vp+\vq)^2} .
\]
For $d=2$ this expression reproduces the previous result \Eqref{xpm}.
Note also that in $d=2$ the equations \eqref{qdse2} for the dressing functions
are finite. (QCD in $d=2$ spatial dimensions is super renormalizable).


\section{Energy density and variational equations}
\label{sec:en}
The calculation of the vacuum expectation values of the Hamiltonian~$\vev{H}$
proceeds completely analogously to the $d=3$ case performed in Refs.~\cite{Campagnari:2015zsa,Campagnari:2018flz}.
For the energy density $e=\vev{H}/(\Nc V)$ one finds in $d=2$ the following contributions:
The single-particle Hamiltonian [first term in the second line of \Eqref{hh1}] yields
\begin{equation}\label{eD}
e_{\mathrm{D}} = \int \dfr[2]{q} \tr \bigl[ (\valpha\cdot\vq + \beta m) \, Q(\vq)\bigr]
- g C_F \int \dfr[2]{q} \dfr[2]{\ell} \, D_{ij}(\vq+\vl) \tr\bigl[\alpha_i Q(\vq) \bar\Gamma_j(\vq,\vl) Q(-\vl) \bigr] ,
\end{equation}
while the kinetic energy of the gluon [the term $J_A^{-1}\Pi_i^a J_A \Pi_i^a$ in \Eqref{hh1}] gives
\begin{equation}\label{eE}
e_E^q = -\frac{C_F}{8} \int\dfr[2]{q} \dfr[2]{\ell} \, t_{ij}(\vq+\vl)
\begin{aligned}[t]
\tr \bigl\{ & \bar\Gamma_{0,i}(\vq,-\vl) Q(\vl) \bar\Gamma_j(\vl,-\vq) Q(\vq) \\
&- Q_0(\vq) \bar\Gamma_{0,i}(\vq,-\vl) Q(\vl) Q_0(\vl)\bar\Gamma_{0,j}(\vl,-\vq) Q(\vq)\bigr\} .
\end{aligned}
\end{equation}
From the Coulomb term we find
\begin{equation}\label{eC}
e_\mathrm{C}^{qq} \simeq - g^2 \frac{C_F}{2} \int \dfr[2]{q} \dfr[2]{\ell} \: F(\vq-\vl) 
\tr\bigl\{ \bigl[Q(\vl)-\tfrac12 Q_0(\vl)\bigr] \bigl[Q(\vq)-\tfrac12 Q_0(\vq)\bigr] - \tfrac14 \bigr\} ,
\end{equation}
where $F$ is the expectation value of the Coulomb kernel $F_A$ [\Eqref{coulkernel}].
Formally, these are exactly the same expressions as in $d=3$ (except for the
momentum integration measure). However, the differences arise now when taking
the traces of the Dirac matrices. In $d=2$ the trace of the unit matrix yields a
factor 2 instead of 4, and the trace of one $\beta$ and an even number of $\alpha_i$
does not vanish like in $d=3$. In particular, we have
\[
\tr[\beta \alpha_i \alpha_j] = 2 \mkern2mu \I* \eps_{ij} .
\]
Although this last expression could in principle make a difference in the calculation,
it turns out that \emph{in the bare vertex-approximation}, where we replace the full
quark-gluon vertices $\bar\Gamma$ [\Eqref{fqgv}] by the bare one $\bar\Gamma_0$ [\Eqref{bqgv}],
this does not matter. The reason is that the 
matrix $\beta$ in the fermion propagator $Q$ always occurs between two projectors~\Eqref{proj}
and leads to expressions of the form
\[
(1+\valpha\cdot\uvp) \beta (1+\valpha\cdot\uvp) = \beta (1-\valpha\cdot\uvp) (1+\valpha\cdot\uvp) = 0 .
\]
Therefore, in the bare-vertex approximation we recover for the energy densities
Eqs.~\eqref{eD}--\eqref{eC} the very same
expressions found in Ref.~\cite{Campagnari:2018flz} apart from the dimension of the
momentum integrals and an overall factor $1/2$. The explicit expressions read in $d=2$
\begin{subequations}\label{loe1}
\begin{align}
\label{loe1a}
e^{}_\mathrm{D} &= 
- 2 \int\dfr[2]{q} \, \frac{\abs{\vq} \, A_q}{\Delta_q} \nonumber \\
&{}\qquad + g^2 C_F \int \dfr[2]{q} \dfr[2]{\ell} \, 
\frac{X_-(\vq,\vl) \, V(\vq,\vl)(A_q \, A_\ell + B_q \, B_\ell) + X_+(\vq,\vl) \, W(\vq,\vl)(A_q \, B_\ell + B_q \, A_\ell)}{\Delta_q \, \Delta_\ell \, \Omega(\vq+\vl)} , \\
\label{loe1b}
e_E^q &= \frac{g^2 C_F}{2} \int \dfr[2]{q} \dfr[2]{\ell} \, \frac{A_q A_\ell}{\Delta_q \Delta_l} \bigl[ X_-(\vq,\vl) \, V^2(\vq,\vl) + X_+(\vq,\vl) \, W^2(\vq,\vl) \bigr] , \\
\label{loe1c}
e_\mathrm{C}^{qq} &= - g^2 \frac{C_F}{4} \int \dfr[2]{q} \dfr[2]{\ell} \: F(\vq-\vl)
\frac{4 B_q \, B_\ell + \uvq\cdot\uvl \bigl[ A_q(2-A_q)-B_q^2\bigr] \bigl[ A_\ell(2-A_\ell)-
B_\ell^2\bigr]}{\Delta_q \Delta_\ell} ,
\end{align}
\end{subequations}
where we have introduced the abbreviation
\[
\Delta_p = A^2_p+B^2_p .
\]

Since the energy density \Eqref{loe1} differs from its three-dimensional
counterpart only by an overall factor and the DSEs \eqref{qdse2} have the same
form in $d=2$ and $d=3$ (except for the explicit expression for $X_\pm$), it is
clear that the variational equations differ from their $d=3$ counterparts only
in the number of dimensions of the momentum integrals, while all numeric factors are exactly
the same. Minimization of the vacuum energy density \Eqref{loe1} with respect to the
vector kernels $V$ and $W$ yields
\begin{align*}
V(\vp,\vq) &= - \frac{1 + s_p s_q}{\Omega(\vp+\vq) + \abs{\vp} \frac{1-s_p^2+2s_p s_q}{1+s_p^2} + \abs{\vq} \frac{1-s_q^2+2s_p s_q}{1+s_q^2}} ,\\
W(\vp,\vq) &= - \frac{ s_p - s_q }{\Omega(\vp+\vq) + \abs{\vp} \frac{1-s_p^2-2s_p s_q}{1+s_p^2} + \abs{\vq} \frac{1-s_q^2-2s_p s_q}{1+s_q^2}} .
\end{align*}
We recall here that the vector kernel $W$ vanishes when $s_p=0$ and is therefore
of purely non-perturbative nature, since the scalar kernel $s_p$ vanishes at any
order in perturbation theory for a vanishing current quark mass.

The variation of the energy density \Eqref{loe1} with respect to the scalar kernel $s_p$ yields the
gap equation
\begin{equation}\label{x1}
\begin{aligned}[b]
\abs\vp s_p
={}& \frac{g^2 C_F}{2} \int\dfr[2]{q} \frac{F(\vp-\vq)}{1+s_q^2} \bigl[ s_q (1-s_p^2) - \uvp\cdot\uvq \, s_p (1-s_q^2) \bigr] \\
&+ \frac{g^2 C_F}{2} \int\dfr[2]{q} \, \frac{s_p}{1+s_q^2} \bigl[ X_-(\vp,\vq) \, V^2(\vp,\vq) + X_+(\vp,\vq) \, W^2(\vp,\vq) \bigr] \\
&-\frac{g^2 C_F}{2}\int\dfr[2]{q} \frac{1}{(1+s_q^2) \Omega(\vp+\vq)} \\
&\quad \times \biggl\{ X_-(\vp,\vq) \, V(\vp,\vq) \bigl[ (1-s_p^2)s_q - 2s_p) \bigr] + X_+(\vp,\vq) \, W(\vp,\vq) \bigl[1-s_p^2-2s_ps_q \bigr] \\
&\qqquad + \frac{\abs\vp}{1+s_p^2} \Bigl[ X_-(\vp,\vq) \, V^2(\vp,\vq) \bigl[ s_p(s_p^2-3)+s_q(1-3s_p^2) \bigr] \\
&\hspace{8em} + X_+(\vp,\vq) \, W^2(\vp,\vq)\bigl[ s_p(s_p^2-3) - s_q(1-3s_p^2) \bigr] \Bigr] \\
&\qqquad+ \frac{\abs{\vq}}{1+s_q^2} \Bigl[ X_-(\vp,\vq) \, V^2(\vp,\vq) \bigl[ (1-s_p^2)s_q - s_p(1-s_q^2) \bigr] \\
&\hspace{8em} - X_+(\vp,\vq) \, W^2(\vp,\vq) \bigl[(1-s_p^2)s_q+s_p(1-s_q^2) \bigr] \Bigr] \biggr\} .
\end{aligned}
\end{equation}
We stress again that, like the DSEs \eqref{qdse2}, also this gap equation has the same
form as its $d=3$ counterpart, however, with $X_\pm$ now given by \Eqref{xpm}.

In $d=3$ we had found \cite{Campagnari:2016wlt} that the addition of the vector
kernel~$W$ makes the gap equation UV finite; there the Coulomb integral
[first integral on the right-hand side of \Eqref{x1}] is logarithmically
divergent, and the integrals involving $V$ and $W$ are (separately) both linearly
and logarithmically divergent. The linear divergence stemming from $W$ cancels
the one stemming from $V$, and the three logarithmic divergences cancel altogether.
In $d=2$ all integrals have one superficial degree of divergence lower than in $d=3$,
and quite remarkably the same cancellation
of divergences happens also in this case, although the tensorial structures $X_\pm$
[\Eqref{xpm}] look quite differently. Here the Coulomb term is UV finite, and the
integrals involving $V$ and $W$ are separately logarithmically divergent but
in the gap equation \eqref{x1} these logarithmic divergences cancel.
As in $d=3$ we find also in this case that the
addition of the vector kernel $W$ makes the gap equation finite; a summary of
the UV divergent contributions is given in Table~\ref{tab:div}.
In fact, we have checked that the leading-order divergence of the gap equation \eqref{x1}
cancels in any number of dimensions once both $V$ and $W$ are considered.

In $d=3$ the vector kernel $W$ was crucial also to ensure multiplicative renormalizability
of the quark propagator \cite{Campagnari:2018flz,Campagnari:2019zso}; this is not
the case here, since the DSEs~\eqref{qdse2} are UV finite.

\begin{table}
\begin{tabular}{l @{\quad} >{$\displaystyle}l<{$} @{\qquad} >{$\displaystyle}l<{$}}
\toprule
& d=3 & d=2 \\
\midrule
Coulomb term        & - \frac{g^2 C_F}{(4\pi)^2} \, s_p \, \abs{\vp} \, \frac{8}{3} \ln\Lambda & \text{finite} \\[3ex]
Terms involving $V$ & \frac{g^2 C_F}{(4\pi)^2} \, s_p \biggl[ -2\Lambda + \abs{\vp} \ln\Lambda \biggl(-\frac{2}{3}+\frac{4}{1+s_p^2}\biggr) \biggr] 
& - \frac{g^2 C_F}{(4\pi)^2} \, s_p \ln\Lambda \\[4ex]
Terms involving $W$ & \frac{g^2 C_F}{(4\pi)^2} \, s_p \biggl[ 2\Lambda + \abs{\vp} \ln\Lambda \biggl(\frac{10}{3}-\frac{4}{1+s_p^2}\biggr) \biggr]
& \frac{g^2 C_F}{(4\pi)^2} \, s_p \ln\Lambda \\
\bottomrule
\end{tabular}
\caption{Comparison of the $d=3$ and $d=2$ UV divergences of the gap equation~\eqref{x1}
stemming from the Coulomb term, the kernel~$V$, and the kernel~$W$.}
\label{tab:div}
\end{table}


\section{Numerical results}
\label{sec:num}
In $d=2$ spatial dimensions the squared coupling constant~$g^2$ has the dimension
of energy, and we express all dimensionful quantities in terms of~$g^2$.
The colour Coulomb potential can be assumed in the form
\[
g^2 F(\vp) = \frac{g^2}{\vp^2} + \frac{2\pi\sigma^{}_{\mathrm{C}}}{\abs{\vp}^3}
\]
which consists of the perturbative part ($\propto1/\vp^2$) and the linearly
rising, confining part. For the gluon propagator \Eqref{gluonprop} we use the
Gribov formula \cite{Gribov:1977wm}
\begin{equation}\label{gribov}
\Omega(\vp) = \sqrt{\vp^2 + \frac{m_A^4}{\vp^2}} ,
\end{equation}
which excellently fits the lattice data in $d=3$ \cite{Burgio:2008jr}.
The infrared analysis of the ghost propagator DSE reveals a relation between the
Gribov mass $m_A$ and the Coulomb string tension $\sigma_{\mathrm{C}}$. When the
angular approximation is used one finds \cite{Feuchter:2007mq}
\[
m_A^2 = \frac{5 N_{\mathrm{c}}}{12} \, \sigma^{}_{\mathrm{C}} ,
\]
while abandoning the angular approximation one obtains \cite{Schleifenbaum:2006bq}
\[
m_A^2 = 4 N_{\mathrm{c}} \biggl(\frac{\Gamma(3/4)}{\Gamma(1/4)}\biggr)^{\!\!2} \sigma^{}_{\mathrm{C}} .
\]
The two values are numerically very close to each other. The Coulomb string
tension~$\sigma_{\mathrm{C}}$ is an upper bound for the Wilson string tension~$\sigma$ \cite{Zwanziger:2002sh},
and in three spatial dimension we have
$\sigma_{\mathrm{C}}\simeq 4 \sigma$ \cite{Greensite:2015nea}.
We have no reliable data for the ratio $\sigma_{\mathrm{C}}/\sigma$
in $d=2$. Since we are interested mostly in a qualitative analysis we choose
$\sigma_{\mathrm{C}} \approx \sigma$.
For the Wilson string tension we take the value \cite{Karabali:1998yq,Bringoltz:2006zg}
\[
\sigma = g^4 \frac{\Nc^2-1}{8\pi} .
\]

\begin{figure}
\centering
\includegraphics[width=.45\linewidth]{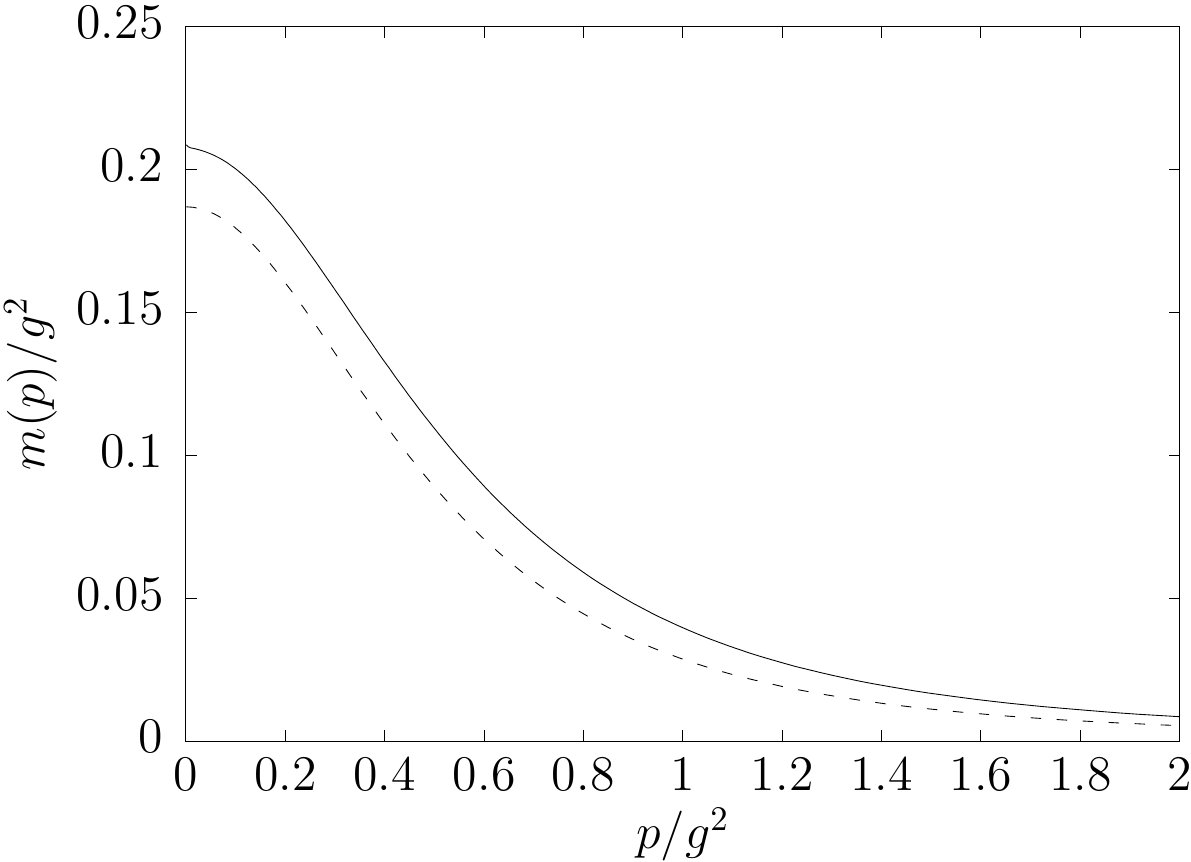}\hfill
\includegraphics[width=.45\linewidth]{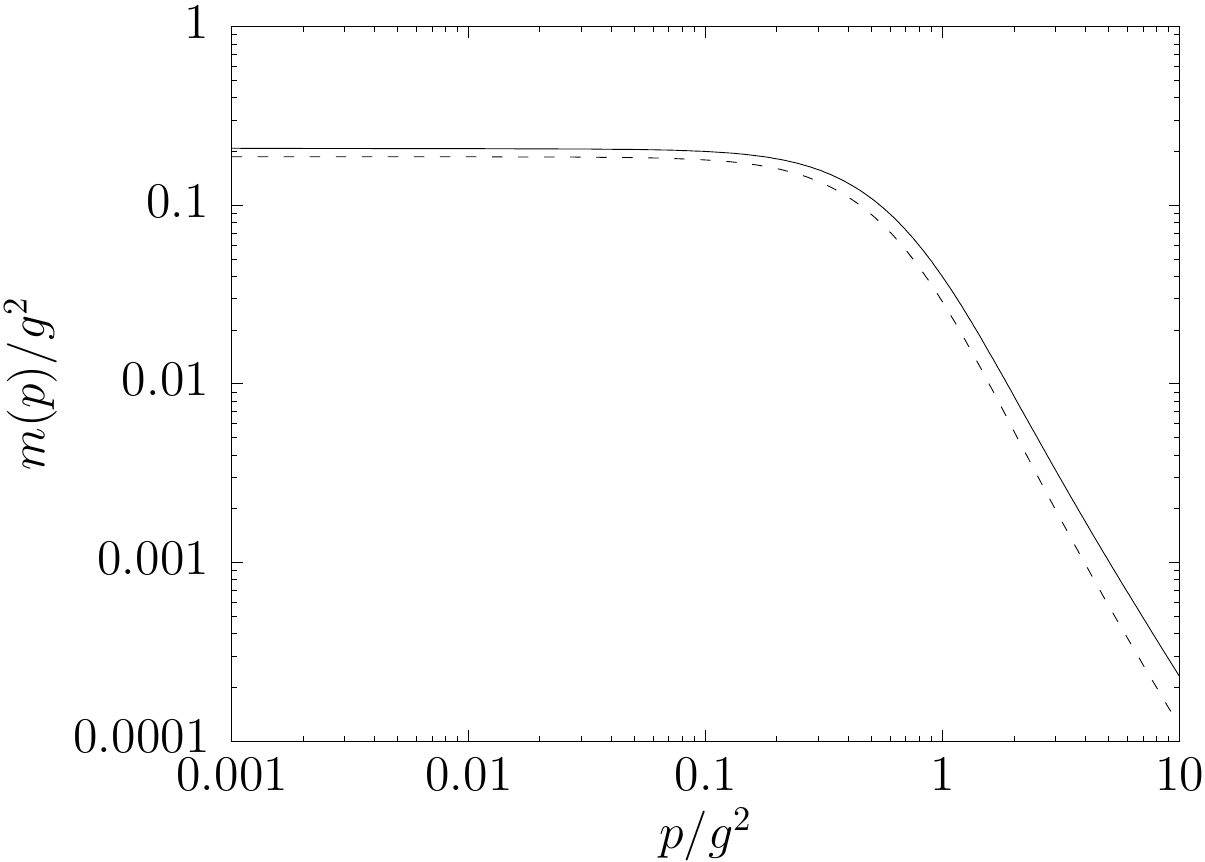}
\caption{Results (left: linear plot, right: logarithmic plot) for the pseudo-mass
function $m_p$ in units of $g^2$ with the colour Coulomb potential alone (dashed
line) and with the coupling to the transverse gluons included (continuous line).}
\label{fig:res}
\end{figure}

For numerical stability it is convenient to reformulate the gap equation~\eqref{x1} in
terms of the pseudo-mass function
\[
m(p) = \frac{2 p s_p}{1-s^2_p} .
\]
The resulting gap equation can be found in Refs.~\cite{Vastag:2015qjd,Campagnari:2016wlt}.
The results of the numerical solution of this equation are shown in Fig.~\ref{fig:res}. Like in the three-dimensional case,
the main contribution to the dynamical mass generation comes from the colour
Coulomb potential [first line in \Eqref{x1}]. The inclusion of the coupling to
the transverse gluons only slightly increases the mass function.

\section{Conclusions}
\label{sec:conc}
In this paper we have investigated the dynamical generation of mass in QCD in $d=2$ spatial
dimensions within the Hamiltonian approach in Coulomb gauge. Somewhat surprisingly, despite
the fundamental differences in the representation of the Lorentz group most results obtained
in $d=3$ hold also in $d=2$. In particular, the inclusion of the non-perturbative vector
kernel $W$ in the bare quark-gluon vertex $\bar\Gamma_0$ [\Eqref{bqgv}] (in addition to the
leading kernel $V$, which exists also in perturbation theory) makes the gap equation UV finite
as in $d=3$. Furthermore, also like in $d=3$, the coupling of the quarks to the spatial gluons
only slightly increases the dynamical mass generation. Like in $d=3$ this effect is absolutely dominated
by the colour Coulomb potential \Eqref{coulkernel}, which results through the elimination of the
temporal gluons $A_0$ in the Hamiltonian approach and, in fact, represents the instantaneous
part of the propagator $\vev{A_0 A_0}$.


\begin{acknowledgments}
This work was supported by the Deutsche Forschungsgemeinschaft (DFG) under contract
No.~DFG-Re856/10-1.
\end{acknowledgments}


\bibliographystyle{h-physrev5}
\bibliography{biblio-spires}

\end{document}